\documentclass[aps,prl,twocolumn,superscriptaddress,showpacs,floatfix,nofootinbib]{revtex4-1}
\usepackage[english]{babel}
\usepackage[T1]{fontenc}
\usepackage[colorlinks=true,linkcolor=blue,citecolor=blue,urlcolor=blue]{hyperref}
\usepackage{bm,bbm,amssymb,amsmath}
\usepackage{graphicx}
\usepackage{isotope}
\usepackage{physics}
\usepackage{multirow,booktabs}

\begin{document}

\title{Properties of Nuclei up to $\boldsymbol{A=16}$ using Local Chiral Interactions}

\author{D. Lonardoni}
\affiliation{Facility for Rare Isotope Beams, Michigan State University, East Lansing, Michigan 48824, USA}
\affiliation{Theoretical Division, Los Alamos National Laboratory, Los Alamos, New Mexico 87545, USA}
\author{J. Carlson}
\author{S. Gandolfi}
\affiliation{Theoretical Division, Los Alamos National Laboratory, Los Alamos, New Mexico 87545, USA}

\author{J.~E.~Lynn}
\affiliation{Institut f\"ur Kernphysik, Technische Universit\"at Darmstadt, 64289 Darmstadt, Germany}
\affiliation{ExtreMe Matter Institute EMMI, GSI Helmholtzzentrum f\"ur Schwerionenforschung GmbH, 64291 Darmstadt, Germany}

\author{K.~E.~Schmidt}
\affiliation{Department of Physics, Arizona State University, Tempe, Arizona 85287, USA}

\author{A.~Schwenk}
\affiliation{Institut f\"ur Kernphysik, Technische Universit\"at Darmstadt, 64289 Darmstadt, Germany}
\affiliation{ExtreMe Matter Institute EMMI, GSI Helmholtzzentrum f\"ur Schwerionenforschung GmbH, 64291 Darmstadt, Germany}
\affiliation{Max-Planck-Institut f\"ur Kernphysik, Saupfercheckweg 1, 69117 Heidelberg, Germany}

\author{X.~B. Wang}
\affiliation{School of Science, Huzhou University, Huzhou 313000, China}

\begin{abstract}
We report accurate quantum Monte Carlo calculations of nuclei
up to $A=16$ based on local chiral two- and three-nucleon interactions
up to next-to-next-to-leading order. We examine the theoretical
uncertainties associated with the chiral expansion and the
cutoff in the theory, as well as the associated
operator choices in the three-nucleon interactions.
While in light nuclei the cutoff variation and systematic uncertainties
are rather small, in \isotope[16]{O} these can be significant
for large coordinate-space cutoffs.
Overall, we show that chiral interactions constructed to reproduce
properties of very light systems and nucleon-nucleon scattering
give an excellent description of binding energies, charge radii,
and form factors for all these nuclei, including open-shell systems
in $A=6$ and 12.
\end{abstract}

\maketitle

\emph{Introduction.---}Predicting the emergence of nuclear properties 
and structure from first
principles is a formidable task.
An important open question is whether it is possible to
describe nuclei and their global properties, e.g.,
binding energies and radii, from microscopic
nuclear Hamiltonians constructed to reproduce only few-body observables,
while simultaneously predicting properties of matter, including the
equation of state and the properties of neutron stars.
Despite advanced efforts, definitive answers are not yet
available~\cite{Barrett:2013nh,Hagen:2013nca,Binder:2013xaa,
Lahde:2013uqa,Carlson:2015,Hebeler:2015hla,Ekstrom:2015,
Hergert:2015awm,Simonis:2017dny}.

Several properties of nuclei up to \isotope[12]{C} have been
successfully described using the phenomenological 
Argonne $v_{18}$ (AV18) nucleon-nucleon ($NN$) potential combined
with Illinois models for the three-body interactions~\cite{Carlson:2015}.
Unfortunately these phenomenological models 
have at least two main limitations.
They do not provide a systematic way to improve the interactions or to
estimate theoretical uncertainties.
In addition, they provide a too soft equation
of state of neutron matter~\cite{Sarsa:2003,Maris:2013}, with
the consequence that the predicted structure of neutron stars
is not compatible with recent observations of two solar-mass
stars~\cite{Demorest:2010,Antoniadis:2013}.

The Argonne-Illinois models have been constructed
to be nearly local: The dominant parts of the interaction
do not depend on the momenta of the two interacting nucleons but only
on their relative distance, spin, and isospin.
This construction was motivated by the ease of employing
such potentials in continuum quantum Monte Carlo (QMC) methods, such as
the Green's function Monte Carlo (GFMC) and
auxiliary field diffusion Monte Carlo (AFDMC) methods. The advantage
of QMC methods is that they can be used to solve accurately and
nonperturbatively the many-body problem without requiring the use of
softer Hamiltonians. The GFMC and AFDMC methods can be 
successfully used only for nearly local Hamiltonians
because of the sign problem~\cite{Carlson:2015}.

In the last two decades, chiral effective field theory (EFT)
has paved the way to the development of nuclear interactions and
currents in a systematic way~\cite{Epelbaum:2008ga,Machleidt:2011zz}.
Chiral EFT expands the nuclear interaction in the ratio of a small scale
(e.g., the pion mass or a typical momentum scale in the nucleus) to a
hard scale (the chiral breakdown scale).
Such an expansion provides several advantages over the traditional
approach, including the ability to improve the interaction order by
order, means to estimate theoretical uncertainties, and the fact that
many-body forces and currents are predicted consistently.
The long-range pion-exchange contributions are determined
by pion-nucleon couplings, while the short-range contributions (given by
so-called low-energy constants) are fit to reproduce
experimental data. Usually, chiral EFT interactions are formulated
in momentum space, but recently Gezerlis {\it et al.} demonstrated
a way to produce equivalent local formulations of chiral $NN$ 
interactions up to next-to-next-to-leading-order 
(N$^2$LO)~\cite{Gezerlis:2013ipa,Gezerlis:2014}. Consistent
three-body forces were
constructed in Refs.~\cite{Tews:2016,Lynn:2016,Lynn:2017fxg}, as
well as chiral interactions with explicit Delta degrees of
freedom~\cite{Piarulli:2015,Piarulli:2016vel,Piarulli:2018,
Ekstrom:2017}.

To solve for the ground state of nuclei, we use the AFDMC
method with local chiral interactions that have been determined from
fits to $NN$ scattering, the alpha particle binding energy, and
$n$--$\alpha$ scattering~\cite{Lynn:2016,Lynn:2017fxg}.
This method has previously been used to determine the
properties of homogeneous and inhomogeneous neutron 
matter~\cite{Gandolfi:2011,Gandolfi:2012,Buraczynski:2016,Buraczynski:2017}, 
and nuclear matter and finite nuclei using simplified
potentials~\cite{Gandolfi:2014}.

In this Letter we present several new important achievements:
(i) the first application of the AFDMC method to calculate properties of
nuclei using chiral Hamiltonians at N$^2$LO, including
three-body forces, (ii) a systematic investigation of the chiral
expansion, including truncation error estimates, in selected nuclei from
$A=3$ to $A=16$, and (iii) an investigation of the cutoff dependence and
the use of different three-body operators for
$A\ge6$.

\emph{Hamiltonian and AFDMC method.---}The Hamiltonian is of the form
\begin{align}
H=-\frac{\hbar^2}{2m}\sum_i \nabla_i^2+\sum_{i<j}v_{ij}+\sum_{i<j<k}V_{ijk} \,,
\end{align}
where the two-body interaction $v_{ij}$ also includes Coulomb and other
electromagnetic effects. 
The two-body potentials $v_{ij}$ and three-body potential $V_{ijk}$
are as in Refs.~\cite{Gezerlis:2013ipa,Gezerlis:2014,Tews:2016,Lynn:2016,
Lynn:2017fxg}.
The general form of the variational state is the following:
\begin{align}
|\Psi\rangle=[F_C+F_2+F_3]|\Phi\rangle_{J,T} \,,
\end{align}
where $F_C$ accounts for all the spin- and isospin-independent correlations, 
and $F_2$ and $F_3$ are linear in spin- and isospin-pair two- and three-body correlations
as described in Ref.~\cite{Carlson:2015}.

The term $|\Phi\rangle$ is taken to be a shell-model-like state
with total angular momentum $J$ and total isospin $T$.
Its wave function
consists of a sum of Slater determinants $D$ constructed using
single-particle orbitals:
\begin{align} 
\langle RS |\Phi\rangle_{J,T} = \sum_n c_n\left(\sum D\big\{\phi_\alpha(\vb{r}_i,s_i)\big\}\right)_{J,T} \,,
\end{align}
where $\vb{r}_i$ are the spatial coordinates of the nucleons and $s_i$ 
represent their spins.
Each single particle orbital $\phi_\alpha$ consists of a radial function $\varphi(r)$
coupled to the spin and isospin states.
The determinants are coupled with Clebsch-Gordan coefficients to total
$J$ and $T$, and the $c_n$ are variational 
parameters multiplying different components having the same quantum numbers.
The radial functions $\varphi(r)$ are obtained by solving for the
eigenfunctions of a Wood-Saxon well,
and all parameters are chosen by minimizing the variational
energy as described in Ref.~\cite{Sorella:2001}.
In order to improve $|\Phi\rangle$, we include single particle orbitals
up to the $sd$ shell.

A complete description of the AFDMC method using two-body interactions
is given in Refs.~\cite{Schmidt:1999,Carlson:2015}. 
Here we describe how three-body interactions are included.
The main limitation of the AFDMC method is that 
the standard Hubbard-Stratonovich transformation used to
propagate the wave function in imaginary time can only be applied
to potentials that are quadratic in spin and isospin
operators. The three-body coordinate-space dependence
is straightforward to include, as are several important terms in the 
three-body interaction that depend on
the spin and isospin of two nucleons at a time. 
Terms depending on the spin and isospin of all three nucleons
are included in an effective way in the propagation,
and then fully accounted for in the final results.
In practice, we determine a Hamiltonian $H'$ that 
mimics the full Hamiltonian, as discussed in the following,
and then we calculate as a perturbation the difference 
$\langle H'-H\rangle$. This procedure goes beyond
the standard normal ordering that averages the
dependence of the third nucleon's position, spin, and isospin.

The chiral three-body interactions at N$^2$LO contain terms that can be
organized as
\begin{align}
V=V_a^{2\pi,P}+V_c^{2\pi,P}+V^{2\pi,S}+V_D+V_E \,.
\end{align}
The first, second, and third terms correspond to the two-pion exchange
diagrams in $P$ and $S$ waves [Eqs.~(A.1b), (A.1c) and (A.1a),
respectively, of Ref.~\cite{Lynn:2017fxg}].
The subscripts $a$ and $c$ refer to the fact that these contributions
can be written in terms of an anticommutator or commutator,
respectively.
We can rewrite $V_{a,c}^{2\pi,P}$ by separating it into long-,
intermediate-, and short-range parts:
\begin{align}
V_{a,c}^{2\pi,P}=V_{a,c}^{XX}+V_{a,c}^{X\delta}+V_{a,c}^{\delta\delta} \,,
\end{align}
where $X$ and $\delta$ refer to the $X_{ij}(\vb{r})$ and
$\delta_{R_{3N}}(\vb{r})$
functions defined in Ref.~\cite{Lynn:2017fxg}.
$V_D$ contains an intermediate-range one-pion-exchange-contact
interaction [Eq.~(24b) of Ref.~\cite{Lynn:2017fxg}], while $V_E$ contains
a short-range term. 
In this work, we consider two alternative forms for $V_E$:
namely, $V_{E\tau}$ and $V_{E\mathbbm1}$ 
[Eqs.~(26a) and (26b), respectively, of Ref.~\cite{Lynn:2017fxg}].
They differ in the operator structure, according to the Fierz-rearrangement 
freedom in the selection of local contact operators in the three-body sector
up to N$^2$LO~\cite{Epelbaum:2002}. $E\tau$ refers to
the choice of the two-body operator $\bm\tau_i\cdot\bm\tau_j$, 
while $E\mathbbm1$ to the choice of the identity operator $\mathbbm1$.

The terms $V_a^{2\pi,P}$, $V^{2\pi,S}$, $V_D$, and $V_E$ are 
purely quadratic in spin and isospin operators, and can be included
exactly in the AFDMC propagator.
The term $V_c^{2\pi,P}$ contains instead explicit cubic
spin and isospin operators. 
These terms cannot be fully included in the AFDMC propagation; however,
their expectation value can be calculated.
We determine the Hamiltonian $H'$ that can be fully propagated as
\begin{align}
H'=H-V_c^{2\pi,P}+\alpha_1 V_a^{XX}+\alpha_2 V_D+\alpha_3 V_E \,.
\end{align}
The three constants $\alpha_i$ are adjusted in order to have
\begin{align}
\langle V_c^{XX}\rangle &\approx\langle\alpha_1 V_a^{XX}\rangle \,, \nonumber \\
\langle V_c^{X\delta}\rangle &\approx\langle\alpha_2 V_D\rangle \,, \nonumber \\
\langle V_c^{\delta\delta}\rangle &\approx\langle\alpha_3 V_E\rangle \,,
\end{align}
where the identifications are suggested by the similar ranges and
functional forms.
The average $\langle\,\cdots\rangle$ indicates an average 
over the propagated wave function.
Once the ground state $\Psi$ of $H'$ is calculated with the AFDMC
method, the expectation value of the Hamiltonian $H$ is given by
\begin{align}
\langle H\rangle&\approx\langle\Psi|H'|\Psi\rangle-\langle\Psi|H'-H|\Psi\rangle \,,
\end{align}
where the last quantity in the previous expression is 
evaluated perturbatively.
Adjusting the constants $\alpha_i$ in such a way that 
the correction is small
suggests that the correction is perturbative.
The same estimate is used in GFMC calculations to determine
the small contributions from nonlocal terms that are 
present in the AV18 potential, and in that case the 
difference $v_8'-v_{18}$ is calculated as a
perturbation~\cite{Pudliner:1997}.

In order to test the technique described above, we first determined the
optimal parameters $\alpha_i$ for a given system, then changed their
values by up to 10\%, and verified that the final result of
$\langle H\rangle$ is nearly independent of such a variation. For example, 
for $^{16}$O
such a variation changes $\langle H'-H\rangle$ from $\approx 1$ to
$\approx 15\,\rm MeV$, but the final estimate of the ground-state energy is 
within $2\,\rm MeV$. In addition, we benchmarked the energies of $A=3$ and
$A=4$ nuclei using the AFDMC method, by comparing with the GFMC results
of Refs.~\cite{Lynn:2016,Lynn:2017fxg}, where the three-body interactions
are included fully in the propagation and found very good agreement
within a few percent.
Note that in many other approaches the three-body force is replaced
by an effective two-body interaction (this is achieved by normal
ordering) neglecting the residual three-body
term~\cite{Hagen:2007ew,Hagen:2012fb}.
However, this approximation has only been benchmarked for softer
interactions~\cite{Roth:2012,Dyhdalo:2017}.

The AFDMC method used here is limited by a sign
problem~\cite{Schmidt:1999,Gandolfi:2014}.
The sign problem is initially suppressed by evolving the wave function
in imaginary time using the constrained-path approximation, where
the configurations are constrained to have positive real overlap with the
trial function, as described in Ref.~\cite{Zhang:2003}. After an initial 
equilibration of the configurations using the constrained-path
approximation, the constraint is removed, and then the evolution in
imaginary-time is performed until the sign-problem dominates and the
variance of the results becomes severely large. The final (statistical) error
strongly depends on the quality of the trial wave function.
We have made several tests to check the results and the dependence on
the initial trial wave function, and have concluded that the systematic 
uncertainties due to releasing the constraint give results correct
to $\sim5\%$ for \isotope[16]{O}. Initial attempts to improve the wave function 
for \isotope[16]{O} show a lowering of the energy by about $4-5\,\rm MeV$, but
since the computational cost is much higher and statistical errors are 
similar to this difference, we leave more detailed studies
to future work.

\emph{Results.---}We consider chiral Hamiltonians at leading-order (LO), 
next-to-leading order (NLO), and N$^2$LO. In this way, following
Ref.~\cite{Epelbaum:2015}, we can assign theoretical uncertainties to
observables coming from the truncation of the chiral expansion.
Uncertainties for an observable $X$ are estimated as 
$\Delta
X^{\text{N}^2\text{LO}}=\max(Q^4\times|X^\text{LO}|,
Q^2\times|X^\text{NLO}-X^\text{LO}|,
Q\times|X^{\text{N}^2\text{LO}}-X^\text{NLO}|)$,
where we take $Q=m_\pi/\Lambda_b$ with $\Lambda_b=600\,\rm MeV$
(see Ref.~\cite{Lynn:2017fxg} for a detailed discussion on uncertainty 
estimates with local chiral interactions).

In Table~\ref{tab:res} we report the AFDMC results for the ground-state
energies and charge radii for nuclei with $A\ge 6$ at N$^2$LO.
In particular, we used the two different cutoffs $R_0=1.0$ and
$R_0=1.2\,\rm fm$ (approximately corresponding to cutoffs in momentum space 
of $500$ and $400\,\rm MeV$~\cite{Lynn:2017fxg}, 
note, however also Ref.~\cite{Hoppe:2017}), 
and two of the three available $V_E$ interactions
constructed in Ref.~\cite{Lynn:2016}.
We find that, starting from local chiral Hamiltonians
fit to $NN$ scattering data~\cite{Gezerlis:2014} 
and three-body interactions fit to light
nuclei~\cite{Lynn:2016,Lynn:2017fxg}, energies and 
radii for nuclei up to $A=16$ are qualitatively well reproduced. 
In particular, we find that the two cutoffs employed here, $R_0=1.0$
and $R_0=1.2\,\rm fm$, reproduce experimental binding energies and charge radii
up to $A=6$ within a few percent. An exception is for the charge radius of 
\isotope[6]{Li} that is slightly underestimated for both cutoffs.
Sizably different is the case of the softer interaction $(R_0=1.2\,\rm fm)$ 
for larger systems, which can significantly overbind \isotope[16]{O},
resulting in a very compact system. In this case the theoretical
uncertainties on the energy are large, dominated by the severe 
overbinding at LO $(\approx -1110\,\rm MeV)$.

We also find that the two different forms ($E\tau$, $E\mathbbm{1}$)
for the three-body interaction give similar results (agreeing within the
EFT uncertainty) for nuclei up to $A=16$.
This suggests that the theoretical uncertainties coming from the
truncation of the chiral expansion are sufficient to account for the
violation of the Fierz rearrangement~\cite{Lynn:2016,Huth:2017wzw}.

\begin{table}[tb]
\centering
\caption[]{Ground-state energies and charge radii for $A=6$, $12$, and $16$ obtained
for the N$^2$LO interactions with different cutoffs $R_0$ and different
three-body interactions. The first uncertainty listed is statistical
while the second is systematic. Experimental results are also shown.}
\begin{tabular}{lccc}
\hline\hline
Nucleus&$V_{E},R_0\,(\rm fm)$ &$E_{\rm AFDMC}\,(\rm MeV)$ &$r_\text{ch}\,(\rm fm)$\\
\hline
\isotope[6]{He} & $E\tau$,      $1.0$ & $-28.4(4)(2.0)$ & $1.99(4)(8)$ \\
   	  	        & $E\mathbbm1$, $1.0$ & $-28.2(5)(1.9)$ & $2.01(4)(7)$ \\
                & $E\tau$,      $1.2$ & $-29.3(1)(1.8)$ & $1.92(4)(8)$ \\
                & Exp                 & $-29.3        $ & $2.068(11)$~\cite{Mueller:2007} \\[0.2cm]
\isotope[6]{Li} & $E\tau$,      $1.0$ & $-31.5(5)(2.3)$ & $2.33(4)(10)$ \\
                & $E\mathbbm1$, $1.0$ & $-30.7(4)(2.1)$ & $2.33(4)(10)$ \\
                & $E\tau$,      $1.2$ & $-32.3(3)(1.7)$ & $2.24(4)(6) $ \\
                & Exp                 & $-32.0        $ & $2.589(39)$~\cite{Nortershauser:2011} \\[0.2cm]
\isotope[12]{C} & $E\tau$,      $1.0$ & $-78(3)(9)    $ & $2.48(4)(18)$ \\
                & Exp                 & $-92.2        $ & $2.471(6)$~\cite{Sick:1982} \\[0.2cm]
\isotope[16]{O} & $E\tau$,      $1.0$ & $-117(5)(16)  $ & $2.71(5)(13)$ \\
                & $E\mathbbm1$, $1.0$ & $-115(6)(15)  $ & $2.72(5)(11)$ \\
                & $E\tau$,      $1.2$ & $-263(26)(56) $ & $2.17(5)(11)$ \\
                & Exp                 & $-127.6       $ & $2.730(25)$~\cite{Sick:1970} \\
\hline\hline
\end{tabular}
\label{tab:res}
\end{table}

\begin{figure}[t]
\includegraphics[width=\linewidth]{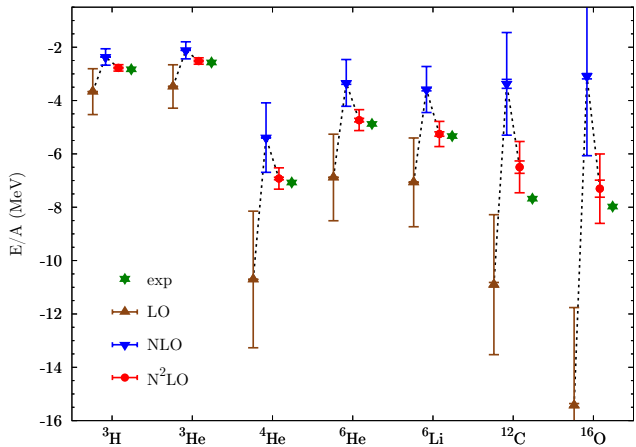}
\caption[]{Ground-state energies per nucleon for
$3\le A\le16$ up to N$^2$LO~$(E\tau)$ with the $R_0=1.0\,\rm fm$ cutoff.
Smaller error bars (indistinguishable from the
symbols up to $A=6$) indicate the statistical Monte Carlo uncertainty,
while larger error bars are the uncertainties from the truncation of
the chiral expansion.}
\label{fig:ene}
\end{figure}

In Fig.~\ref{fig:ene} we present the ground-state energies per nucleon
of selected nuclei with $3\le A\le16$, calculated at LO, NLO, and
N$^2$LO~$(E\tau)$ with the cutoff $R_0=1.0\,\rm fm$.
The error bars are estimated by including the statistical uncertainties
given by the AFDMC calculations as well as the error given by
the truncation of the chiral expansion.
The ground-state energies per nucleon are in agreement
with experimental data up to $A=6$, while for \isotope[12]{C} and
\isotope[16]{O} the energies are somewhat underpredicted.
The uncertainties are reasonably small, dominated by the truncation error.

\begin{figure}[b]
\includegraphics[width=\linewidth]{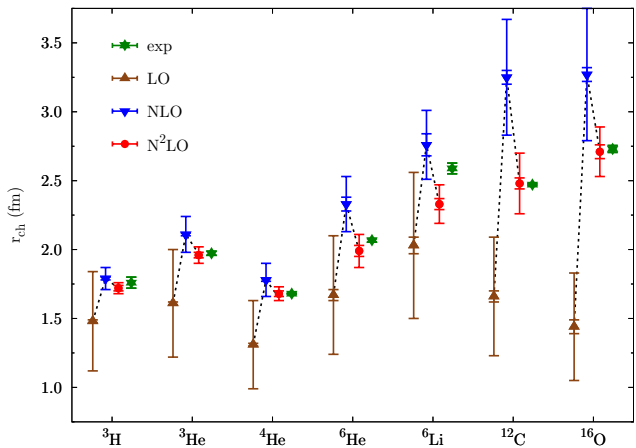}
\caption[]{Charge radii for $3\le A\le16$ up to
N$^2$LO~$(E\tau)$ with the $R_0=1.0\,\rm fm$ cutoff.
Error bars are as in Fig.~\ref{fig:ene}.}
\label{fig:rch}
\end{figure}

In Fig.~\ref{fig:rch} we compare the charge radii calculated at LO, NLO,
and N$^2$LO~$(E\tau)$ with the $R_0=1.0\,\rm fm$ cutoff to experimental data. 
These results show that a qualitative description of binding energies and charge
radii is possible starting from Hamiltonians constructed using only
few-body data. We note, however, that the radius of \isotope[6]{Li} is slightly smaller
than the experimental measurement. It is interesting to note that the
charge radius of \isotope[6]{Li}
calculated with the GFMC method employing the AV18 and
Illinois VII~(IL7) three-body interactions is also
underestimated~\cite{Carlson:2015}.

\begin{figure}[b]
\includegraphics[width=\linewidth]{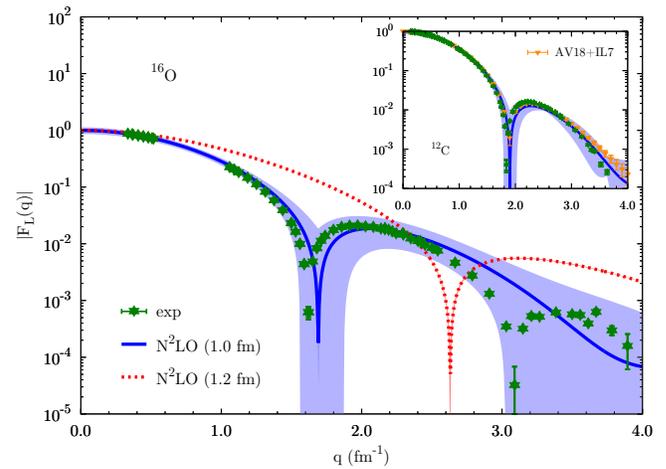}
\caption[]{Charge form factor for \isotope[16]{O} at N$^2$LO for $R_0=1.0$ and 
$1.2\,\rm fm$ compared to experimental data~\cite{Sick:1970,Schuetz:1975,Sick:1975}.
For $R_0=1.0\,\rm fm$, both $E\tau$ and $E\mathbbm1$ three-body operators give consistent
results. The shaded area indicates the statistical Monte Carlo uncertainty combined
with the (dominant) uncertainty from the truncation of the chiral expansion. 
For \isotope[12]{C}, AFDMC results are shown in the inset for 
$R_0=1.0\,\rm fm$ versus experimental data from Ref.~\cite{Devries:1987} and the 
GFMC results employing the AV18+IL7 potentials~\cite{Lovato:2013}.}
\label{fig:ff}
\end{figure}

We show in Fig.~\ref{fig:ff} the charge form factors of \isotope[12]{C} and
\isotope[16]{O} compared to
experimental data. The \isotope[12]{C} form factor is also compared
to previous GFMC calculations with the AV18+IL7 potentials.
Our form factor calculations have been 
performed using one-body charge operators only. Two-body operators
are expected to give small contributions only at momenta larger than 
$\approx 500\,\rm MeV$~\cite{Lovato:2013,Mihaila:2000}, as they basically include
relativistic corrections.
It is interesting to compare the curves given by the two
different cutoffs. In the figure, the result obtained using $R_0=1.0\,\rm fm$ at N$^2$LO~$(E\tau)$
(solid blue line) includes the uncertainty from the truncation of the chiral expansion
(shaded blue area). The agreement with experimental data is very good. 
For $R_0=1.2\,\rm fm$ at N$^2$LO~$(E\tau)$ (dotted red line), the radius is too small
and the first diffraction minimum occurs at a significantly higher
momentum than experimentally observed, consistent with the overbinding
obtained for this interaction.

\begin{figure}[t]
\includegraphics[width=\linewidth]{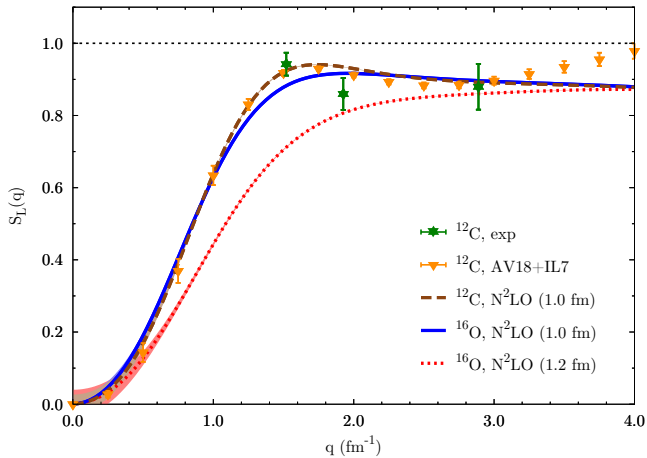}
\caption[]{Coulomb sum rules for \isotope[12]{C} and \isotope[16]{O} at N$^2$LO~$(E\tau)$
compared to experimental data as derived in
Ref.~\cite{Lovato:2016}. Results are shown for $R_0=1.0$ and $1.2\,\rm fm$ for 
\isotope[16]{O}, and for $R_0=1.0\,\rm fm$ for \isotope[12]{C}. Shaded areas indicate
only the statistical Monte Carlo uncertainty. In addition, we compare our results to
GFMC calculations for \isotope[12]{C} with AV18+IL7~\cite{Lovato:2013}.}
\label{fig:sl}
\end{figure}

Finally, in Fig.~\ref{fig:sl} we present the Coulomb sum rules for \isotope[12]{C} and \isotope[16]{O}.
The AFDMC result for \isotope[12]{C} is compatible both with the available experimental
data as extracted in Ref.~\cite{Lovato:2016} and
with the GFMC result for AV18+IL7~\cite{Lovato:2013}. 
The differences between the AFDMC and GFMC results at high momentum are due to two-body currents, 
fully implemented to date only in the GFMC calculations. 
For \isotope[16]{O}, the result for the harder interaction with $R_0=1.0\,\rm fm$
is very close to that of \isotope[12]{C}, and is compatible with the findings
of Ref.~\cite{Lonardoni:2017} for the AV18+UIX potential. The softer
interaction with $R_0=1.2\,\rm fm$ produces instead a significantly
different result, as for the charge form factor.

\emph{Summary.---}We have performed QMC calculations of selected nuclei up to $A=16$ using 
local chiral interactions at LO, NLO, and N$^2$LO for different cutoffs and three-body
interactions. We conclude that these Hamiltonians, constructed
only from $NN$ data and properties of few-body nuclei,
can give a good description of ground-state properties of nuclei up to $A=16$, 
including binding energies, charge radii, form factors, and Coulomb sum rules. 
This is true in particular for the harder interaction considered here, corresponding to 
coordinate-space cutoff $R_0=1.0\,\rm fm$. 
For the larger cutoff $R_0=1.2\,\rm fm$, we find in \isotope[16]{O} a strong 
dependence of the energy uncertainty coming from the truncation of the chiral expansion,
and a large overbinding and compactness. The latter two could be a consequence of the large $c_D$ 
coupling in the $E\tau$ parametrization of the three-body force~\cite{Lynn:2016}, 
resulting in a sizable attractive contribution not present in the hard interaction $(c_D=0)$.
More detailed analysis to further investigate this behavior will be performed in future works.

We thank I.~Tews, A.~Lovato, A.~Roggero, C.~Petrie, and K.~Hebeler for
many valuable discussions.
The work of D.L. was was supported by the U.S. Department of Energy,
Office of Science, Office of Nuclear Physics, under the FRIB Theory
Alliance Grant Contract No. DE-SC0013617 titled ``FRIB Theory Center - A
path for the science at FRIB,'' and by the NUCLEI SciDAC program.
The work of J.C. and S.G. was supported by the NUCLEI SciDAC program,
by the U.S. Department of Energy, Office of Science, Office of Nuclear
Physics, under contract No. DE-AC52-06NA25396, and by the LDRD program
at LANL.
The work of J.E.L. and A.S. was supported by the ERC Grant No.~307986
STRONGINT and the BMBF under Contract No.~05P15RDFN1.
K.E.S. was supported by the National Science Foundation Grant No.
PHY-1404405.
X.B.W. thanks the hospitality and financial support of LANL and the
National Natural Science Foundation of China under Grant No. U1732138,
No. 11505056, and No. 11605054.
Computational resources have been provided by Los Alamos Open
Supercomputing via the Institutional Computing (IC) program, by the
National Energy Research Scientific Computing Center (NERSC), which is
supported by the U.S. Department of Energy, Office of Science, under
contract DE-AC02-05CH11231, and by the Lichtenberg high performance
computer of the TU Darmstadt.

\end{document}